\documentclass[letterpaper]{amsart}
\pdfoutput=1
\newtheorem{theorem}{Theorem}[section]

\newtheorem{corollary}[theorem]{Corollary}

\theoremstyle{definition}

\theoremstyle{remark}
\newtheorem{remark}[theorem]{Remark}
\usepackage{hyperref}
\usepackage{graphicx}
\usepackage{color}
\usepackage{amsmath}
\usepackage{amsfonts}
\usepackage{amssymb}
\usepackage{epsfig}
\usepackage{rotating}
\usepackage{graphics}
\usepackage{mathrsfs}
\newcommand{\be}{\begin{equation}}

\newcommand{\ee}{\end{equation}}

\newcommand{\R}{{\bf R }}

\newcommand{\ba}{\begin{array}}

\newcommand{\ea}{\end{array}}

\newcommand{\beq}{\begin{eqnarray}}

\newcommand{\eeq}{\end{eqnarray}}

\newtheorem{lm}{lemma}

\newtheorem{thee}{theorem}

\newtheorem{proo}{proposition}

\newtheorem{co}{corollary}

\newtheorem{rem}{remark}

\newtheorem{deff}{definition}

\newcommand{\bd}{\begin{deff}}

\newcommand{\ed}{\end{deff}}

\newcommand{\bl}{\begin{lm}}

\newcommand{\el}{\end{lm}}

\newcommand{\bp}{\begin{proo}}

\newcommand{\ep}{\end{proo}}

\newcommand{\bt}{\begin{thee}}

\newcommand{\et}{\end{thee}}

\newcommand{\bc}{\begin{co}}

\newcommand{\ec}{\end{co}}

\newcommand{\brm}{\begin{rem}}

\newcommand{\erm}{\end{rem}}

\newcommand{\der}{{\rm d}}

\usepackage{t1enc}
\def\frak{\mathfrak}

\newcommand{\newc}{\newcommand}

\renewcommand{\exp}{\operatorname{exp}}

\let\ccdot\cdot
\def\cdot{\hbox to 2.5pt{\hss$\ccdot$\hss}}

\newc{\aR}{\mbox{\boldmath{$ R$}}}
\newc{\aS}{\mbox{\boldmath{$ S$}}}
\newc{\aT}{\mbox{\boldmath{$ T$}}}
\newc{\aW}{\mbox{\boldmath{$ W$}}}

\newc{\aK}{\mbox{\boldmath{$ K$}}}
\newc{\aL}{\mbox{\boldmath{$ L$}}}
\newcommand{\bbC}{\mathbb{C}}
\newcommand{\bbN}{\mathbb{N}}

\usepackage{amssymb}
\usepackage{amscd}

\newcommand{\bma}{\begin{pmatrix}}
\newcommand{\ema}{\end{pmatrix}}

\newc{\obstrn}[2]{B^{#1}_{#2}}

\newcommand{\rpl}                         % +) or <+
{\mbox{$
\begin{picture}(12.7,8)(-.5,-1)
\put(0,0.2){$+$}
\put(4.2,2.8){\oval(8,8)[r]}
\end{picture}$}}

\newcommand{\lpl}                         % (+ or +>
{\mbox{$
\begin{picture}(12.7,8)(-.5,-1)
\put(2,0.2){$+$}
\put(6.2,2.8){\oval(8,8)[l]}
\end{picture}$}}

\usepackage{ifthen}

\newc{\tensor}[1]{#1}
\newc{\Mvariable}[1]{\mbox{#1}}
\newc{\down}[1]{{}_{#1}}
\newc{\up}[1]{{}^{#1}}

\newc{\JulyStrut}{\rule{0mm}{6mm}}
\newc{\midtenPan}{\mbox{\sf S}}
\newc{\midten}{\mbox{\sf T}}
\newc{\midtenEi}{\mbox{\sf U}}
\newc{\ATen}{\mbox{\sf E}}
\newc{\BTen}{\mbox{\sf F}}
\newc{\CTen}{\mbox{\sf G}}

\def\sideremark#1{\ifvmode\leavevmode\fi\vadjust{\vbox to0pt{\vss% the remark
 \hbox to 0pt{\hskip\hsize\hskip1em%                          will appear only
 \vbox{\hsize3cm\tiny\raggedright\pretolerance10000%          on the side
 \noindent #1\hfill}\hss}\vbox to8pt{\vfil}\vss}}}%
\numberwithin{equation}{section}

\newcounter{romenumi}
\newcommand{\labelromenumi}{(\roman{romenumi})}

\begin{document}
\title[Poincare-Einstein approach to Conformal Cyclic Cosmology]{Poincare-Einstein approach to Penrose's Conformal Cyclic Cosmology}
\dedicatory{Dedicated to Robin C. Graham for the occasion of his 65th birthday}
\vskip 1.truecm
\author{Pawe\l~ Nurowski} \address{Centrum Fizyki Teoretycznej, Polska Akademia Nauk, Al. Lotnik\'ow 32/46, 02-668 Warszawa, Poland}
\email{nurowski@cft.edu.pl}
\thanks{The research was funded from the Norwegian Financial Mechanism 2014-2021 with project registration number 2019/34/H/ST1/00636.}
\begin{abstract}
We consider two consecutive eons in Penrose's Conformal Cyclic Cosmology and study how the matter content of the past eon determines the matter content of the present eon by means of the reciprocity hypothesis of Roger Penrose.

We assume that the only matter content in the final stages of the past eon
is a spherical wave described by Einstein's equations with a pure radiation
energy momentum tensor and with a cosmological constant. 

Using the Poincare-Einstein type of expansion to determine the metric in the past eon, applying the reciprocity hypothesis to get the metric in the present eon, and using the Einstein equations in the present eon to interpret its matter content, we show that the single spherical wave from the previous eon in the new eon splits into three portions of
radiation: the two spherical waves, one which is a damped continuation
from the previous eon, the other is focusing in the new eon as it encountered a mirror at the Big Bang surface, and in addition a lump of scattered radiation described by the statistical physics.
  \end{abstract}

\date{\today}
\maketitle
%*************
%\vspace{-1truecm}
%\tableofcontents
\newcommand{\bbS}{\mathbb{S}}
\newcommand{\bbR}{\mathbb{R}}
\newcommand{\sog}{\mathbf{SO}}
\newcommand{\glg}{\mathbf{GL}}
\newcommand{\slg}{\mathbf{SL}}
\newcommand{\og}{\mathbf{O}}
\newcommand{\soa}{\frak{so}}
\newcommand{\gla}{\frak{gl}}
\newcommand{\sla}{\frak{sl}}
\newcommand{\sua}{\frak{su}}
\newcommand{\dr}{\mathrm{d}}
\newcommand{\sug}{\mathbf{SU}}
\newcommand{\gat}{\tilde{\gamma}}
\newcommand{\Gat}{\tilde{\Gamma}}
\newcommand{\thet}{\tilde{\theta}}
\newcommand{\Thet}{\tilde{T}}
\newcommand{\rt}{\tilde{r}}
\newcommand{\st}{\sqrt{3}}
\newcommand{\kat}{\tilde{\kappa}}
\newcommand{\kz}{{K^{{~}^{\hskip-3.1mm\circ}}}}
\newcommand{\bv}{{\bf v}}
\newcommand{\di}{{\rm div}}
\newcommand{\curl}{{\rm curl}}
\newcommand{\cs}{(M,{\rm T}^{1,0})}
\newcommand{\tn}{{\mathcal N}}
\newcommand{\ten}{{\Upsilon}}
%*************
\noindent
\section{Introduction}
Our common view of the Universe is that it evolves from the \emph{initial singularity} to its present state, when we observe the presence of the \emph{positive} cosmological constant \cite{per,ries,schmi}. This has the remarkable consequence that the Universe will eventually become \emph{asymptotically de Sitter} spacetime. This in particular means that the `end of the Universe' - a hypersurface where all the null geodesics will end bounding the spacetime in the future  - will be \emph{spacelike} \cite{spin}. Moreover, it will be \emph{conformally flat}, pretty much the same as the \emph{initial} boundary of the Universe - the Big Bang \cite{PT1}.

This fundamental facts led Roger Penrose to a cosmology theory proposal termed by him \emph{Conformal Cyclic Cosmology} (CCC). The proposal has a lot of physical motivations, which can be found in \cite{pen} and in numerous lectures of Penrose available even in the public media. In this introduction we will only mention its mathematical background, which will be needed to explain our results.

The main feature of CCC is that it states that its universe consists of \emph{eons}, each being a \emph{time oriented} spacetime, whose \emph{conformal compactifications} have \emph{spacelike} null infinities $\mathscr{I}$. We recall that the future/past null infinity is a boundary $\mathscr{I}^+$/$\mathscr{I}^-$ of spacetime, where all the future/past null geodesics end.  

To avoid confusions in understanding CCC we first emphasize that: 
\begin{itemize} 
 \item CCC says \emph{nothing} about this what is the physics in a given eon when the physical age of it  \emph{is normal};   \emph{normal} here means that the eon is neither \emph{ too young} nor \emph{ too old}.   CCC tells what is going on when an eon is \emph{ either about to die,   or had just been born}.  
   \item In particular, CCC does \emph{not} require that the eons have the same history!   It is Conformal \emph{Cyclic} Cosmology,   and \emph{not}  Conformal \emph{Periodic} Cosmology! 
     \end{itemize}

The framework for CCC was recently shaped by Paul Tod, which can be briefly described as follows (see: \cite{pt}, for details):
\begin{itemize}
\item The universe consists of \emph{eons}, each being a \emph{time oriented} spacetime, whose \emph{conformal compactifications} have \emph{spacelike} $\mathscr{I}^-$ and $\mathscr{I}^+$ .   The \emph{ Weyl tensor} of the 4-metric on each $\mathscr{I}$ \emph{ is zero}.  
\item Eons are ordered, and the \emph{conformal compactifications} of consecutive eons,   say \emph{the past one} and \emph{the present one},   are \emph{glued together}   along $\mathscr{I}^+$ \emph{of the past eon},   and $\mathscr{I}^-$ \emph{of the present eon}.
\item The \emph{manifold} $M$ \emph{of these glued eons is the universe}. The CCC, as formulated by Tod in \cite{pt}, tells what is mathematical structure of the CCC universe in the \emph{neighborhood of matching surfaces of any two consecutive eons}. Each such surface is, in Penroses's imaginative language, a \emph{wound} of the universe. The neighborhood of each wound, consisting of a time portion of the past eon and a time portion of the present eon, is a \emph{bandage region} of the universe - a (conformal) time sandwich in the universe, in which each wound is bandaged to heal the trauma of the (conformal) transition through the Big Crunch/Big Bang.
  \item Each bandage region is equipped with the following \emph{three metrics}, which are \emph{conformally flat} at the wound:  
    \begin{itemize}
    \item a Lorentzian metric $g$ which is regular everywhere, 
    \item a Lorentzian metric $\check{g}$, which represents the physical metric of the \emph{present eon}, and which is \emph{singular} at the wound, 
      \item a Lorentzian metric $\hat{g}$, which represents the physical metric of the \emph{ past eon}, and which \emph{infinitely expands} at the wound.
      \end{itemize}
    \end{itemize}
\begin{itemize}
\item In a bandage region, the \emph{ three metrics} $g$, $\check{g}$ and $\hat{g}$, are \emph{ conformally related} on their overlapping domains.\\
  %\centerline{\includegraphics[scale=0.04]{eons.jpg}}
  \centerline{\includegraphics[scale=0.3]{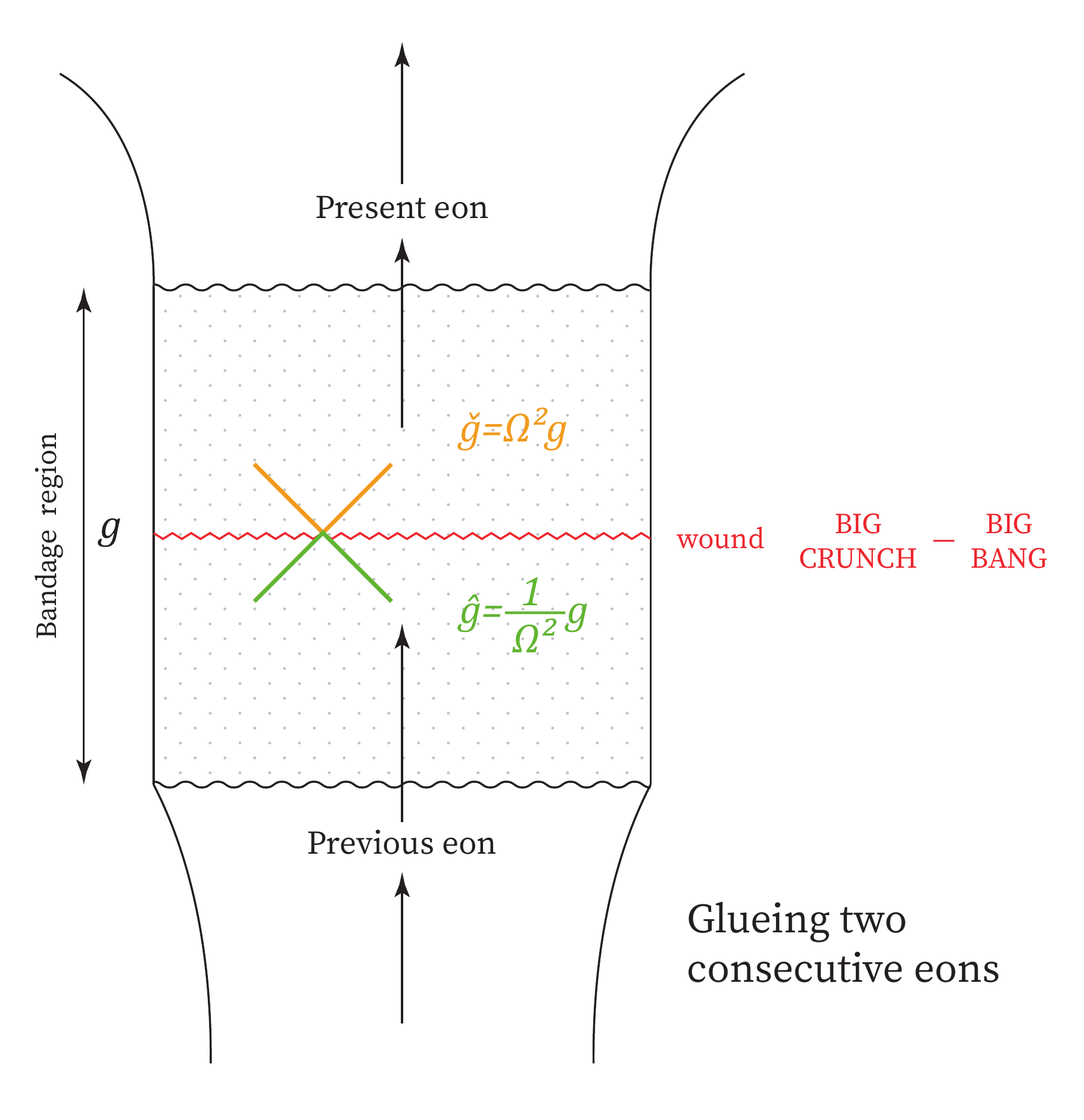}}
\item How to make this relation specific is debatable,   but Penrose proposes that
  \be \check{g}=\Omega^2g \quad\mathrm{and} \quad\hat{g}=\frac{1}{\Omega^2}g,\quad\mathrm{ with}\quad \Omega \to 0\quad\mathrm{on\,\,the\,\, wound}.\label{br}\ee
  This is called \emph{reciprocity hypothesis}. 
  \item The metric $\check{g}$ \emph{ in the present eon} is a \emph{ physical metric there}.   Likewise, the metric $\hat{g}$ \emph{ in the past eon} is a \emph{ physical metric there}.  
  \item Of course, the metric $\check{g}$ \emph{ in the present eon},   and the metric $\hat{g}$ \emph{ in the past eon},   as \emph{ physical spacetime metrics},   \emph{ should satisfy Einstein's equations} in their spacetimes, respectively.
\end{itemize}

To answer a natural \emph{question} on how to make a model of Penrose's bandaged region of two eons,  one needs a function $\Omega$,   \emph{vanishing on some spacelike hypersurface},   and a \emph{regular} Lorentzian 4-metric $g$,    such that \emph{if} $\check{g}=\Omega^2g$ \emph{satisfies Einstein equations} with some physically reasonable energy momentum tensor,   \emph{then} $\hat{g}=\frac{1}{\Omega^2}g$ \emph{also satisfies Einstein equations} with possibly different,  but still physically reasonable energy momentum tensor.

This is a question similar to the question posed and \emph{solved} by \emph{H. Brinkman} \cite{bri}.   In 1925 he asked `\emph{when in a conformal class of metrics there could be two nonisometric Einstein metrics?}'.   Brinkman found all such metrics in dimension \emph{four}.   In every signature.

In CCC the problem is similar. On one hand it seems even simpler: the \emph{same} function $\Omega$ should lead to \emph{two conformally related  but different solutions} $\check{g}=\Omega^2g$ and $\hat{g}=\Omega^{-2}g$ \emph{of Einstein equations},   with a prescribed energy momentum tensor on the $\hat{M}$ part,   and a \emph{reasonable} energy momentum tensor on the other $\check{M}$. This creates a highly overdetermined system of PDEs, which may have no solutions at all, or may create algebraic constraints on the unknown variables, resulting in obtaining the solutions in explicit form (see Section \ref{babar} for an example of such a situation). On the other hand, the problem is not so easy, because the three metrics as in \eqref{br} obtained from the process of solving Einstein equations on both sides of the Big Crunch/Big Bang hypersurface $\mathscr{I}$ \emph{must be conformally flat on} $\mathscr{I}$. This last requirement is automatically satisfied e.g. if one looks for the metrics $g$, $\check{g}$ and $\hat{g}$ which are conformally flat \emph{everywhere} in the bandage region. It is a reasonable simplifying assumption if one wants to test implications of Penrose's CCC proposal in particular situations, such as in the case of metrics conformal to the Robertson-Walker metrics (see Section \ref{cfccc} of the present paper). However to see the implications of the the full CCC setting, the assumption of conformal flatness of the whole bandage region is too strong, and in the general case one is led to consider the full initial value problem on a conformally flat spacelike hypersurface $\mathscr{I}$ for the Einstein equations with a \emph{particular energy momentum tensor} for one of the physical metrics $\hat{g}$ or $\check{g}$. This can be in principle done by considering results of Friedrich \cite{frie1,frie2} for the conformal data on $\mathscr{I}$ applied to the Starobinsky expansion \cite{star} as proposed in the context of CCC  by Tod \cite{PT1}. In the present paper, rather than the Starobinsky expansion we use the Poincare-Einstein type of expansion generalizing for the purpose of CCC the approach to conformal geometry due to Fefferman and Graham \cite{fef}. In our Section \ref{swi}, we first formulate this Poincare-Einstein approach to CCC, and then in Section \ref{purerad} we use our new approach to see what's going on with a spherical wave passing from the past eon to the present one if it is obeying these newly established rules of CCC.

In Section \ref{cfccc} we describe how to produce \emph{conformally flat} bandage models of CCC. Throughout the rest of the article, and in particular in Section \ref{purerad} we \emph{abandon} the conformal flatness everywhere, and create a physically appealing and satisfactory CCC bandage region model which is conformally flat at the Big Bang/Big Crunch hypersurface \emph{only}. 

\section{Conformally flat models in Penrose's Conformal Cyclic Cosmology}\label{cfccc}
To illustrate various CCC bandage region concepts in conformally flat case we will assume in this entire Section that all the three bandage metrics $\hat{g}$, $g$ and $\check{g}$ are conformal to Friedman-Lema\^itre-Robertson-Walker metric
\be g_{test}=-\der t^2+\Omega^2(t)r_0^2~\Big(\der\chi^2+\sin^2\chi\big(\der\theta^2+\sin^2\theta\der\phi^2\big)\Big)\label{flrw}\ee
with (locally) spherical spatial sections $\kappa=1$. Before passing to the CCC details we recall the relevant information about this metric and a perfect fluid.
\subsection{Polytropic perfect fluid in FLRW spherical cosmology}
It is convenient to introduce a {\bf conformal time} $\eta=\int\frac{\der t}{\Omega(t)}$ so that the FLRW metric \eqref{flrw} looks   $$g_{test}=\Omega^2(\eta)\Big(-\der\eta^2+r_0^2\big(\der\chi^2+\sin^2\chi(\der\theta^2+\sin^2\theta\der\phi^2)\big)\Big),$$   i.e. $g_{test}= \Omega^2(\eta) g_{Einst}$. Here
$$ g_{Einst}=-\der\eta^2+r_0^2\big(\der\chi^2+\sin^2\chi(\der\theta^2+\sin^2\theta\der\phi^2)\big)$$
is the Einstein Static Universe metric describing the Einstein Static Universe $M=\bbR\times \bbS^3$, with $\eta$ being Einstein's Universe absolute time, and $(\chi,\theta,\phi)$ being the spherical angles on $\bbS^3$.  

This parametrization is very convenient since taking $u=-\Omega(\eta)\der\eta$, the most general FLRW metric $g$ satisfying \emph{ Einstein's equations}
\be Ric-\frac{1}{2}Rg_{test}=(\mu+p)u\otimes u+pg_{test}\label{efe}\ee with \emph{polytropic equation of state} $p=w\mu$, $w=const$, is given by\\ 
    \centerline{$\Omega(\eta)=\Omega_0\Big(\sin^2\frac{(1+3w)\eta}{2r_0}\Big)^{\frac{1}{1+3w}}$ if $w\neq -\frac{1}{3}$,}   and\\ 
    \centerline{$\Omega(\eta)=\Omega_0\exp(b\eta)$ if $w=-\frac{1}{3}$.}
  \subsection{Bandage region in FLRW framework with perfect fluids without cosmological constants}\label{babar}
We use this \emph{ explicit} solutions to the Einstein field equations \eqref{efe} to create a \emph{ conformally flat everywhere} bandaged region of two consecutive eons via the Penrose-Tod scenario. On doing this we go back to the Penrose-Tod's bandage triple $(\check{g},g,\hat{g})$ and:
      \begin{itemize}   
  \item We take $g$ as $g_{Einst}$, $g=g_{Einst}$; 
  \item We take $\check{g}=g_{test}=\Omega^2(\eta)g_{Einst}$.     This satisfies Einstein's equations with perfect fluid with $\check{p}=\check{w}\check{\mu}$. This means that the conformal scale function $\Omega=\Omega(\eta)$ is:
\be \Omega(\eta)= \left\{ \begin{array}{lll}
  \Omega_0\Big(\sin^2\frac{(1+3\check{w})\eta}{2r_0}\Big)^{\frac{1}{1+3\check{w}}}&\mathrm{ if} &\check{w}\neq -\frac{1}{3}\\
  \Omega_0\exp(b\eta)&\mathrm{ if} &\check{w}=-\frac{1}{3}
\end{array} \right.
\label{eeee}.\ee
\item We take as $\hat{g}=\Omega^{-2}(\eta)g_{Einst}$.\\
   %\centerline{\includegraphics[scale=0.04]{eony.jpg}}
\centerline{\includegraphics[scale=0.5]{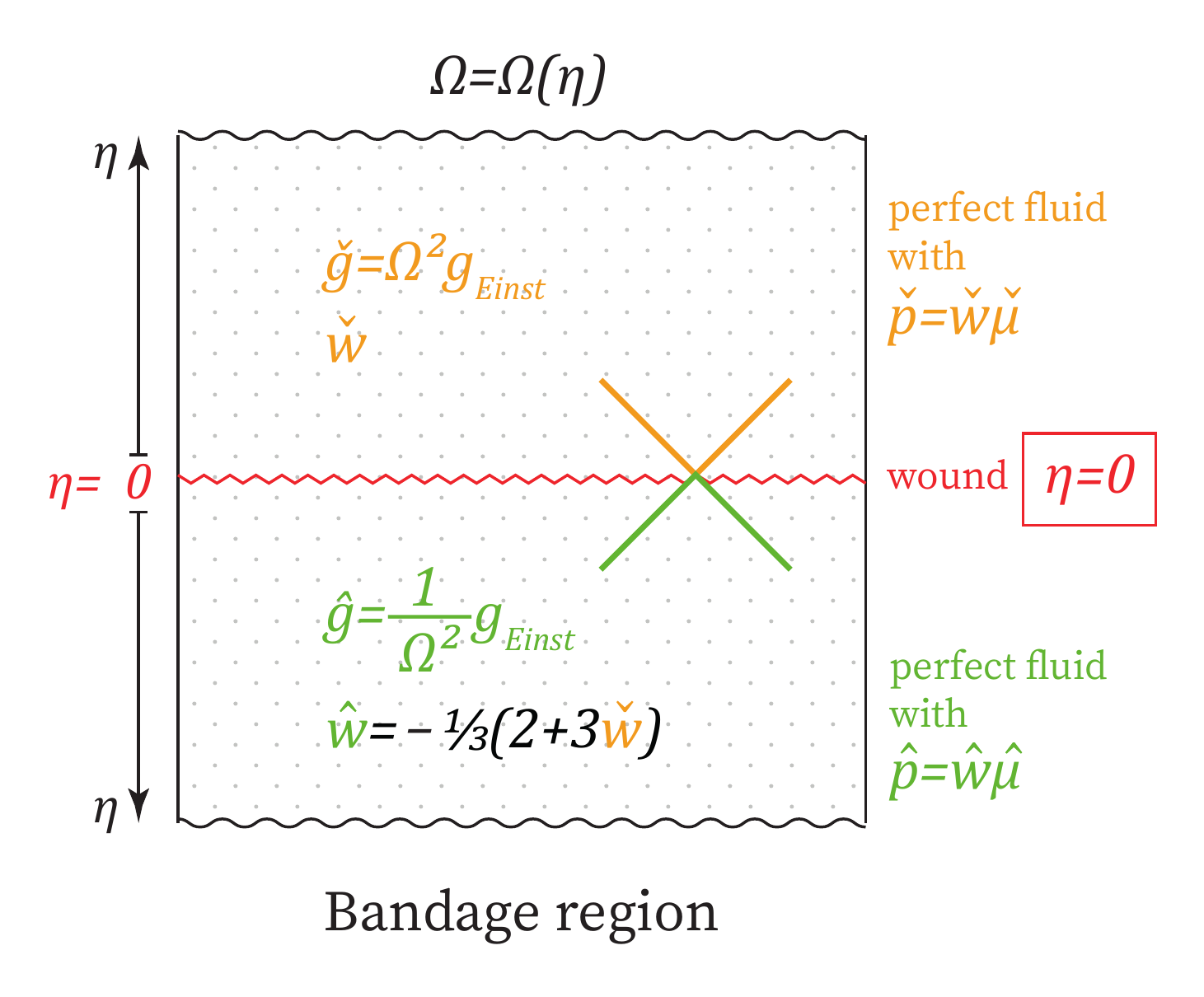}}
\item Since the metric $\check{g}=\Omega^2g$ satisfying postulated Einstein's equations has the scale function $\Omega=\Omega(\eta)$ as in \eqref{eeee}, then its reciprocal $\Omega^{-1}$ has the same functional dependence on $\eta$. Merely the parameter $\check{w}$ will change into $\hat{w}$, with obvious $(1+3\check{w})^{-1}\to -(1+3\hat{w})^{-1}$.
    \item Thus the metric $\hat{g}=\Omega^{-2}g$ satisfies the \emph{ same kind of Einstein's equations}, but now  with $\check{w}$ replaced by $\hat{w}$ such that 
    $\hat{w}=-2/3-\check{w}$.  
    \item In other words $\hat{g}=\Omega^{-2}g_{Einst}=\Omega^{-4}\check{g}$ also satisfies Einstein's equations with perfect fluid,   but with $\hat{p}=\hat{w}\hat{\mu}$. 
      \end{itemize}
      We have the following theorem relating the polytropes in two consecutive eons:\\
      \begin{theorem}
  If $\Omega=\Omega(\eta)$ is such that $\boxed{\check{g}=\Omega^2g_{Einst}}$ \emph{ satisfies Einstein's equations}, with $\check{\Lambda}=0$,   and with \emph{the energy momentum tensor} $\check{T}$ \emph{ of a perfect fluid},   whose pressure $\check{p}$ is proportional to the energy density $\check{\mu}$,   via $\check{p}=\check{w}\check{\mu}$, $\check{w}=const$,   then $\boxed{\hat{g}=\frac{1}{\Omega^2}g_{Einst}}$   \emph{satisfies Einstein's equations}, with $\hat{\Lambda}=0$, and with \emph{the energy momentum tensor} $\hat{T}$ \emph{ of a perfect fluid},   whose pressure $\hat{p}$ and the energy density $\hat{\mu}$ are related by   $\hat{p}=\hat{w}\hat{\mu}$   with\\\centerline{$\boxed{\hat{w}=-\frac{1}{3}(2+3\check{w})}$.}  
  The \emph{ Ricci scalar} of the metric $\hat{g}$ is\\
%Conformal_to_Einstein_Universe_in_-+++_spherical_polytrope_better.nb
%Conformal_to_Einstein_Universe_in_-+++_spherical_polytrope_better_inverse.nb
  $$\hat{R}=
\left\{ \begin{array}{lll}
\frac{3(1-3\hat{w})\Omega_0^2}{r_0^2\big(\sin^6\frac{(1+3\hat{w})\eta}{2r_0}\big)^{\frac{1+\hat{w}}{1+3\hat{w}}}}& \mathrm{if}&\hat{w}\neq-1/3\\
  \\
\frac{6(1+b^2r_0^2)\Omega_0^2\exp(2b\eta)}{r_0^2}&\mathrm{if}&\hat{w}=-1/3
\end{array} \right.,
$$
  so it is \emph{ positive} if $\boxed{-1\leq \hat{w}<1/3}$.
\end{theorem}

      \begin{remark}
        In CCC the consecutive eons should have \emph{spacelike} $\mathscr{I}$s. For this the Ricci scalar $\hat{R}$ of the physical metric $\hat{g}$ \emph{must be positive} at the wound surface (see \cite{spin}, p. 353, or \cite{pt}, p. 8). This together with the dominant energy condition for the fluid in the past eon, $-1\leq\hat{w}\leq 1$, shows that possible values of the $\hat{w}$ parameter is $-1\leq \hat{w}<1/3$.\\
%\centerline{\includegraphics[scale=0.03]{wutowu.jpg}}
\centerline{\includegraphics[scale=0.4]{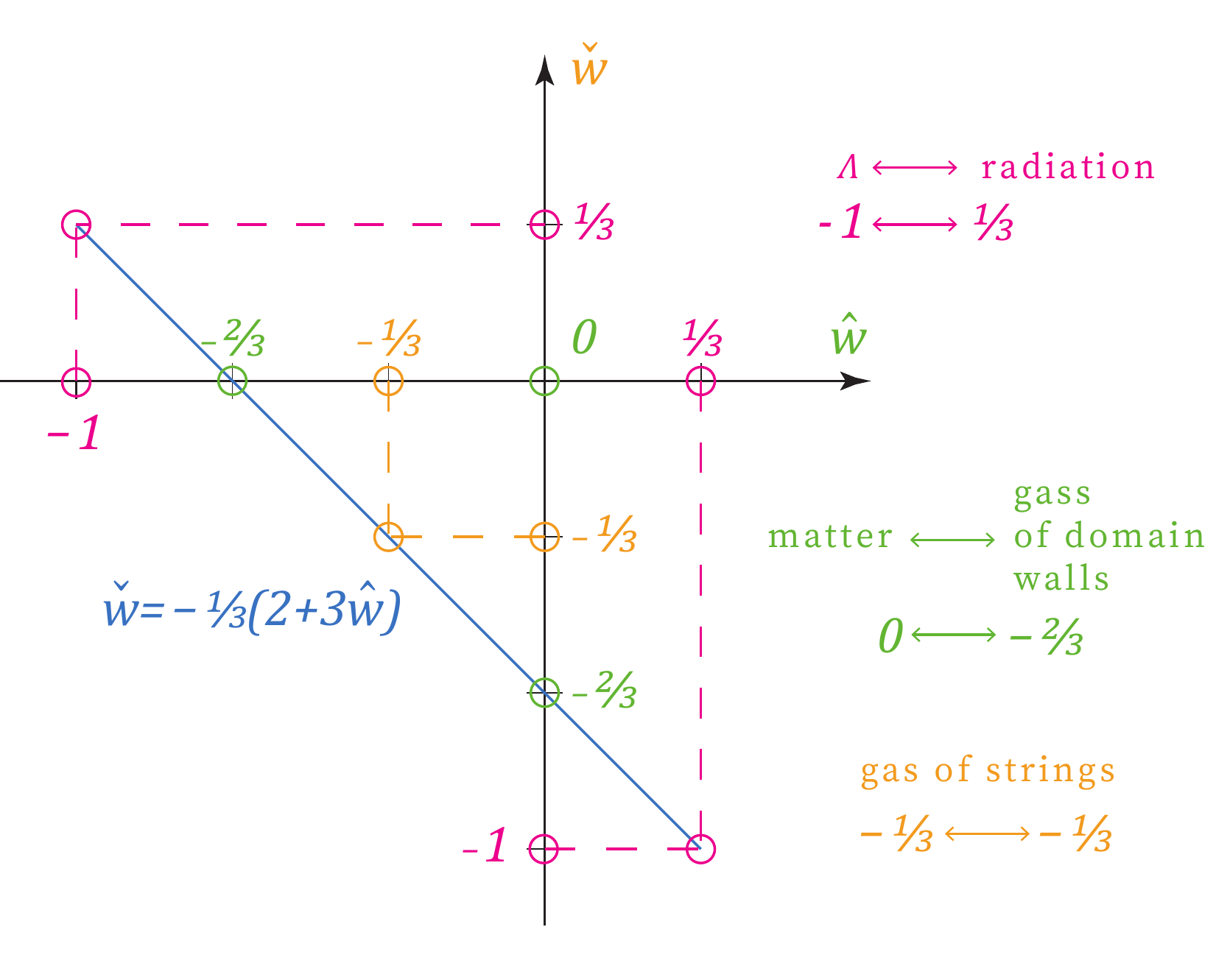}}
        If we have the past eon filled with the perfect fluid with $\hat{p}=\hat{w}\hat{\mu}$ and $\hat{w}\in [-1,1/3[$, then the reciprocity hypothesis transforms it to a present eon filled with the perfect fluid with $\check{w}=-\tfrac13(2+3\hat{w})$. This in particular means that \begin{itemize}\item if \emph{the past eon is the deSitter space}, $\hat{w}=-1$, then \emph{the present eon is filled with radiation}, $\check{w}=1/3$; \item if \emph{the past eon is filled with a gas of domain walls} (see \cite{KT}, p. 219-220), then \emph{the present eon is filled with dust}, $\check{w}=0$; \item if \emph{the past eon is filled with the gas of strings} (see \cite{KT}, p.227), then \emph{the present eon is also filled with gas of strings};
    \item This continues: \emph{the dust in the past eon is transformed into a gas of domain walls in the present eon}. 
    \end{itemize}
    Going along the interval $\check{w}=-\tfrac13(2+3\hat{w})$ on the $(\hat{w},\check{w})$ plane, with $-1\leq\hat{w}<1/3$, we eventually reach the point $(\hat{w},\check{w})=(1/3,-1)$. This point is forbidden however, since for $\hat{w}=1/3$ the $\mathscr{I}^+$ of the past eon becomes \emph{null}. To see how \emph{the radiation} $\hat{w}=1/3$ passes through the Bang surface one needs to consider $\hat{g}$ as a solution of Einstein equations with a \emph{positive cosmological constant} $\hat{\Lambda}$ and with the energy momentum tensor of perfect fluid with $\hat{p}=1/3\hat{\mu}$. We will discuss this point in more detail in Section \ref{radiation}.        \end{remark}
      \begin{remark}
        The transitions of eons' fluids: \emph{cosmological constant}$\longrightarrow$\emph{radiation}, \emph{gas of domain walls}$\longleftrightarrow$\emph{dust} and \emph{gas of strings}$\longleftrightarrow$\emph{gas of strings} was first observed in Ref. \cite{nurmei}. These transformations appeared there as \emph{mysteriously quantized} for only five values of $\hat{w}$, which were integer multiples of the number $1/3$. The above result, which states that \emph{all} values of $\hat{w}$ from the interval $-1\leq \hat{w}<1/3$ are possible, shows that the `quantization' discussed in \cite{nurmei} was merely obtained as a consequence of the assumptions made in \cite{nurmei}, which restricted the search of solutions to only those which were \emph{real analytic} in the time variable $t$. As we have shown here, one does not need to look for solutions of the Einstein equations \eqref{efe} with $p=w\mu$ in the restricted power series form as in \cite{nurmei}. The Einstein equations solve completely in terms of elementary functions! The general solution is not analytic at $t=0$; the solutions which are analytic are those discussed in \cite{nurmei}.  
      \end{remark}
       \subsection{Bandage region in FLRW framework with perfect fluids and cosmological constants}\label{radiation}
       To analyze what happens if the conformally flat past eon satisfies Einstein's equations with a \emph{cosmological constant} and with a \emph{radiative perfect fluid} we return to the general FLRW metric $\hat{g}=\Omega(t)^{-4}\big(-\der t^2+\Omega^2(t)r_0^2g_{\bbS^3}\big)$,
       with the original Friedmann time variable $t$.
    %Conformal_to_Einstein_Universe_in_-+++_spherical_polytrope_past.nb
  Then the condition that $\hat{g}$ satisfies Einstein's equations
$$\hat{Ric}-\tfrac12\hat{R}\hat{g}+\hat{\Lambda}\hat{g}=(1+\hat{w})\hat{\mu}\hat{u}\otimes\hat{u}+\hat{w}\hat{\mu}\hat{g}$$
with $\hat{u}=-\der t/\Omega(t)^2$, $\hat{w}=const$, and the cosmological constant $\hat{\Lambda}$, is equivalent to the following ODE for $\Omega$:
\be 2r_0^2\Omega^3\Omega''=(1+3\hat{w})\Omega^2(1+r_0^2\Omega'{}^2)-(1+\hat{w})\hat{\Lambda} r_0^2.\label{conn}\ee
We want that $\check{g}=\Omega^4\hat{g}$ satisfies the same kind of Eisntein's equations
$$\check{Ric}-\tfrac12\check{R}\check{g}+\check{\Lambda}\check{g}=(1+\check{w})\check{\mu}\check{u}\otimes\check{u}+\check{w}\check{\mu}\check{g}$$
with $\check{u}=-\der t$, and the cosmological constant $\check{\Lambda}$,   and $\check{w}=const$. This gives another second order condition for $\Omega$, which when compared with \eqref{conn} gives: 
%Conformal_to_Einstein_Universe_in_-+++_spherical_polytrope_past_inverse.nb
\be \begin{aligned}3\hat{\Lambda}^2&r_0^2(1+\hat{w})^2+\hat{\Lambda}\check{\Lambda}r_0^2\big(5+2\hat{w}-3\hat{w}^2\big)\Omega^4-\\&9\hat{\Lambda}(1+\hat{w})^2\Omega^2(1+r_0^2\Omega'{}^2)+\check{\Lambda}(1-3\hat{w})^2\Omega^6(1+r_0^2\Omega'{}^2)=0.\end{aligned}\label{conj}\ee
Differentiating this identity with respect to $t$, and eliminating the second derivative of $\Omega$ by means of the ODE \eqref{conn} we get the next identity relating $\Omega$ and $\Omega'$. This reads:
%Conformal_to_Einstein_Universe_in_-+++_spherical_polytrope_past_inverse_dalej1.nb
\be \begin{aligned}
  9\hat{\Lambda}^2&r_0^2(1+\hat{w})^3-\hat{\Lambda}\check{\Lambda}r_0^2(1+\hat{w})(-19+6\hat{w}+9\hat{w}^2)\Omega^4-
  \\&27\hat{\Lambda}(1+\hat{w})^3\Omega^2(1+r_0^2\Omega'{}^2)+\check{\Lambda}(1-3\hat{w})^2(7+3\hat{w})\Omega^6(1+r_0^2\Omega'{}^2)=0,
  \end{aligned}\label{conk}
\ee
where to avoid deSitter solutions on both sides of the bandage region we assumed that $$\Omega'\neq 0.$$
Using this assumption again, after an elimination of the unknown $\Omega'$ from the equations \eqref{conj} and \eqref{conk}, we get the following identity which must be satisfied by the unknown constants $\hat{\Lambda}$, $\check{\Lambda}$ and $\hat{w}$:
    %Conformal_to_Einstein_Universe_in_-+++_spherical_polytrope_past_inverse_dalej1.nb
    $$\check{\Lambda}\hat{\Lambda}(1+\hat{w})(1-3\hat{w})=0.$$
     Thus, a \emph{necessary condition} for both $\Omega$ and $\Omega^{-1}$ to describe the polytropes, is that \emph{either} one of the $\Lambda$s is zero, \emph{ or} $\hat{w}$ is of the `radiation-$\Lambda$' type.

     The case of two $\Lambda$s being zero was considered in the previous section; it follows that the case $\hat{w}=-1$ and both $\Lambda$s nonvanishing leads to the deSitter space on both sides of the bandaged region. \emph{ So here we concentrate on the remaining case when}
     $$\check{\Lambda}\hat{\Lambda}\neq 0\quad\mathrm{and}\quad\hat{w}=1/3.$$

     It follows that in such a case the \emph{Einstein equations imply} that \emph{ remarkably} also $\check{w}=1/3$. This is a generalization of a result from Ref. \cite{pt} stating that if the two $\Lambda$s are nonvanishing and equal and the past eon is filled radiation type perfect fluid, the present eon is also filled with radiation. More explicitly this case can be integrated to the very end, and we have the following theorem.
\begin{theorem}
     %Conformal_to_Einstein_Universe_in_-+++_spherical_polytrope_again_inverse_dalej_w13_solution.nb
      %Conformal_to_Einstein_Universe_in_-+++_spherical_polytrope_again_solution.nb
  The function $\Omega=\Omega(t)$ given by:\\  \centerline{$\Omega^2=\frac{3-3\cosh(2\sqrt{\frac{\check{\Lambda}}{3}}t)-2r_0^2\sqrt{\check{\Lambda}\hat{\Lambda}}\sinh(2\sqrt{\frac{\check{\Lambda}}{3}}t)}{\check{\Lambda}r_0^2}$}  
  has the property that both   $\check{g}=\Omega^2g_{Einst}$   and $\hat{g}=\Omega^{-2}g_{Einst}$   satisfy Einstein's equations with polytropic perfect fluid equation of state, for which $\hat{w}=\check{w}=1/3$   (radiation),   and with the corresponding cosmological constants $\check{\Lambda}$ and $\hat{\Lambda}$. Here $g_{Einst}=-\Omega^{-2}\der t^2+r_0^2g_{\bbS^3}$.
  \end{theorem}

This theorem says that \emph{incoherent radiation happily passes through the wound}.   However, cosmological constants can change from any positive value to any other one. In this respect our result is more general than the observation of Paul Tod from \cite{pt}.

\section{The Poincare-Einstein setting}\label{swi}

In Ref. \cite{pt} Paul Tod proposed that in a given bandage region of the universe, the 4-metric $\hat{g}$ of the past eon $\hat{M}$  should be determined from the initial data, given in terms of the conformal 3-metric $h_0$ on the wound hypersurface, via the \emph{Starobinsky expansion} \cite{star}
\be
\hat{g}=-\der \tau^2+\mathrm{e}^{2\tau}(h_0+\mathrm{e}^{-2\tau}h_2+\mathrm{e}^{-3\tau}h_3+...).\label{pte}\ee
Here the past eon $\hat{M}$ is to be identified with the Cartesian product of $\hat{\mathscr{I}}{}^+$ and the interval $[0,\epsilon[$, $\epsilon>0$, and $h_i$, $i=1,2,3,\dots,$ are \emph{symmetric rank 2 tensors on} $\hat{\mathscr{I}}{}^+$. The tensors $h_i$ defining the physical metric of the past eon $\hat{g}$ as in \eqref{pte} are to be determined by the Einstein equations satisfied by $\hat{g}$ in $\hat{M}$. Paul Tod further (secretly) requires that the metric formula \eqref{pte} makes sense for $\tau\in]-\epsilon',\epsilon'[$, with $0<\epsilon'\leq\epsilon$ and that such a metric solves the same Einstein equations for all  $\tau\in]-\epsilon',\epsilon'[$. This produces a metric $\hat{g}$ in the neighborhood $M=\hat{\mathscr{I}}{}^+\times]-\epsilon',\epsilon'[$ of both time sides of $\hat{\mathscr{I}}{}^+$, i.e. in the full (sufficiently small) bandage region containing $\hat{\mathscr{I}}{}^+$. Once having $\hat{g}$ there is still a  problem in how to define $\check{g}$ in $M$: In order to use  the reciprocity hypothesis one needs the choice of either $\Omega$ or $g$. Tod in \cite{pt} assumes in addition the constancy of the cosmological constant $\hat{\Lambda}$ and finds a PDE for $\Omega$, which can be solved in the cases he considers, giving a unique transition from $\hat{M}$ to the new eon $\check{M}$.

        In our paper we propose a different algorithm for finding the three metrics $\hat{g}$, $g$ and $\check{g}$ defining the geometry in each bandage region.

        This is based on the following observation (mentioned marginally in \cite{pt} by Tod, see the beginning of Section 2 in \cite{pt}): Introduce a new variable $t=-\mathrm{e}^{-\tau}$ and insert it in the Starobinsky expansion \eqref{pte}. This will produce an alternative formula for $\hat{g}$ which is:
        \be \hat{g}=\frac{-\der t^2+h_t}{t^2}=\frac{-\der t^2+h_0+h_2 t^2-h_3t^3+\dots}{t^2}.\label{pte1}\ee
        If such $\hat{g}$ satisfied the Einstein's equations $\hat{Ric}=\hat{\Lambda}\hat{g}$, this would be the Poincar\'e-Einstein metric for the conformal class $[h_0]$ of 3-dimensional metrics defined on $\hat{\mathscr{I}}{}^+$. The power series expansion \eqref{pte1}, precisely in our variable $t$, is well known in the theory of conformal invariants \cite{fef}. In this theory, a result of Fefferman and Graham \cite{fef} guarantees that if the dimension $N$ of the conformal manifold with the conformal structure $[h_0]$ represented by a (pseudo)Riemannian metric $h_0$ is \emph{odd}, and the expansion $h_t=\sum_{i=0}^\infty h_it^i$ contains terms of \emph{even powers of} $t$ (i.e. if $h_{2k+1}=0$ for all $k=1,2,\dots$), then the power series expansion \eqref{pte1} is \emph{uniquely} determined up to infinite order in $t$ by the requirement that $\hat{g}$ satisfies the Einstein equations $\hat{Ric}(\hat{g})=N\hat{g}$ also up to infinite order. Actually, Fefferman and Graham have also a theorem that handles the situation with nonzero \emph{odd terms} in \eqref{pte1}. Since this is more appropriate for the application in the present paper we quote it here\footnote{We adapt the Theorem 4.8 from \cite{fef} to the conformal dimension $N=3$ and the Lorentzian signature $(-,+,+,+)$ of the Poincar\'e-Einstein metric $\hat{g}$}:\\
        \vspace{0.5cm}

        \noindent
        \emph{Fefferman-Graham Theorem} \cite{fef}\\
        Let $h_0$ be a Riemannian metric on a 3-dimensional manifold $\hat{\mathscr{I}}{}^+$ and let $h$ be a smooth symmetric tensor on $\hat{\mathscr{I}}{}^+$ which is \emph{traceless}, $h_0^{ij}h_{ij}=0$, and \emph{divergence free}, $\nabla^jh_{ij}=0$,  w.r.t. $h_0$. Then there exists a 4-dimensional Lorentzian metric $$\hat{g}=\frac{-\der t^2+\sum_{i=0}^\infty h_it^i}{t^2}$$ satisfying the Einstein equations $$\hat{Ric}(\hat{g})=3\hat{g},$$ such that \emph{trace-free-part-of-the-expression}$\Big(\big(\partial_t^3(h_t)\big)_{|t=0}\Big)=h$. These conditions on $h_t$ at order $3$ in $t$ \emph{uniquely} determine $h_t$ to \emph{infinite order} at $t=0$. Moreover, the solution satisfies $h_0^{ij}\big((\partial_t^3h_t)_{|t=0}\big)_{ij}=0$  \\
        \vspace{0.4cm}

        Our discussion above suggests that Tod's formulation of the geometry of bandage CCC regions in terms of $\hat{g}$ being in Starobinsky expansion could also be possible in terms of the Poincar\'e-Einstein expansion \eqref{pte1} of Fefferman and Graham. This new formulation should be advantageous since the Fefferman-Graham \emph{normal form} of the metric $\hat{g}$ distinguishes a particular conformal factor $t$ whose zero defines \emph{conformal infinity} of $\hat{g}$. This conformal factor gives then a natural choice for $\Omega=t$ which painlessly solves the problem of finding missing $\Omega$ out of $\hat{g}$. 

        We therefore propose to give up with the Starobinsky expansion approach and to formulate the problem of finding the three metrics characterizing bandage regions as follows:

        \begin{itemize}
        \item Start with a \emph{Riemannian} conformal class $[h_0]$ on a 3-dimensional manifold $\mathscr{I}$. The class is represented by a Riemannian 3-dimensional metric $h_0$ of your choice. 
        \item Consider the metric \be \hat{g}=\frac{-\der t^2+h_t}{t^2}\label{eineq}\ee in $M=\mathscr{I}\times]-\epsilon,\epsilon[$, $\epsilon>0$, with a power series $h_t=\sum_{i=0}^{\infty}h_it^i$, where for $i=1,2,\dots$, the coefficients $h_i$ are unknown symmetric rank 2 tensors on $\mathscr{I}$; the first term in the expansion $h_0$, corresponding to $i=0$, will then be `your choice' representative $h_0$ of the conformal class $[h_0]$.
        \item Impose the Einstein equations $\hat{Ric}-\tfrac12 \hat{R}\hat{g}+\hat{\Lambda}\hat{g}=\hat{T}$ on the metric \eqref{eineq} and determine the unknown tensors $h_i$, $i=1,2,\dots$ everywhere in $M$, i.e. for all values of $t$ such that $|t|<\epsilon'$, $0<\epsilon'\leq\epsilon$. Here $\hat{T}$ is a \emph{given} energy momentum tensor describing the matter in $\hat{M}$.  
        \item Having determined $\hat{g}$ make the only natural choice and define the scale function $\Omega$ to be $\Omega=t$. Use it and the already determined $h_t$ to define the metric $\check{g}$ to be $\check{g}=t^2(-\der t^2+h_t)$ in $M$. This is an immediate consequence of the reciprocity hypothesis, once $\Omega$ is chosen as above. \item Analyze the matter content of the new eon $\check{M}=\mathscr{I}\times]-\epsilon',0]$, considering the energy momentum tensor $\check{T}:=\check{Ric}-\tfrac12 \check{R}\check{g}$. This is given in the new eon $\check{M}$ via the Einstein equations red back, in the Synge's way, from right to left. 
        \end{itemize}
        
\section{Spherical wave transition from the past to the present eon}\label{purerad}
   In this Section we assume that in the past eon we have a \emph{spherical wave} carrying energy density $\hat{\Phi}$ with speed of light towards its Big Crunch hypersurface $\hat{\mathscr{I}}{}^+$. This in particular means that in $\hat{M}$
%with coordinates $(z,\bar{z},r,t)$, $z\in \bbC$, $r>0$, $t\in\bbR$,
we have a \emph{given} vector field $K^i$, which is \emph{null} with respect to the metrics $\hat{g}=\hat{g}{}_{ij}\der x^i\der x^j$,
$$\hat{g}{}_{ij}K^iK^j=0.$$
It also means that the metric $\hat{g}$ satisfies the Einstein equations 
\be \hat{R}^{ij}=\hat{\Lambda} \hat{g}^{ij}+\hat{\Phi} K^i K^j\label{eincheck}\ee
with this given null vector field $K^i$. We postpone for a while the question \emph{how the spherical symmetry of the wave} with propagation vector $K^i$ \emph{is implemented}. First we adapt our new procedure of obtaining all the three metrics $\hat{g}$, $g$ and $\check{g}$, to the bandaged region of the above defined matter in $\hat{M}$. In this adaptation we will chose the spherical wave in $\hat{M}$ such that the conformal class $[h_0]$ on the Big Crunch hypersurface $\hat{\mathscr{I}}{}^+$ is \emph{flat}. 
\subsection{The ansatz and the model for the past eon}     

We start with a conformal class $[h_0]$ represented by the flat 3-dimensional metric
$$h_0=\frac{2r^2\der z \der\bar{z}}{(1+\frac{z\bar{z}}{2})^2}+\der r^2.$$
Then as $h_t$ in \eqref{eineq} we take the \emph{spherically symmetric} 1-parameter family
$$h_t=\frac{2r^2\big(1+\nu(t,r)\big)\der z \der\bar{z}}{(1+\frac{z\bar{z}}{2})^2}+\big(1+\mu(t,r)\big)\der r^2,$$
where the both unknown functions $\nu=\nu(t,r)$ and $\mu=\mu(t,r)$  admit a \emph{power series expansion} in the variable $t$ such that:
$$\nu(0,r)=0\quad\mathrm{and}\quad\mu(0,r)=0.$$
This obviously satisfies $h_{t=0}=h_0$ and because of our power series assumptions above we have
$$\nu(t,r)=\sum_{i=1}^\infty a_i(r) t^i\quad\mathrm{and}\quad \mu(t,r)=\sum_{i=1}^\infty b_i(r) t^i,$$
with a set of differentiable functions $a_i=a_i(r)$ and $b_i=b_i(r)$ depending on the $r$ variable only.

This leads to the following ansatz for the Poincar\'e-type metric $\hat{g}$ in $\hat{M}$:
\be
\begin{aligned}
\hat{g}=&\hat{g}{}_{ij}\der x^i\der x^j=\frac{\frac{2r^2(1+\nu)}{(1+\frac{z\bar{z}}{2})^2}+(1+\mu)\der r^2-\der t^2}{t^2}=\\
&\frac{2r^2\big(\,1+\sum_{i=1}^\infty a_i(r) t^i\,\big)\der z \der\bar{z}}{t^2(1+\frac{z\bar{z}}{2})^2}\,+\,\frac{1+\sum_{i=1}^\infty b_i(r) t^i}{t^2}\der r^2-\frac{\der t^2}{t^2}.
\end{aligned}
\label{prem}\ee
Our past eon manifold $\hat{M}$ is therefore parameterized by coordinates $(x^i)=(z,\bar{z},r,t)$ with $t>0$, $r>0$ and $z\in\bbC\cup\{\infty\}$.

Now to implement the spherical symmetry of the wave we consider the following null vector field $K$ on $\hat{M}$:
\be K=K^i\partial_{x^i}=\partial_t+\frac{1}{\sqrt{1+\mu}}\partial_r=\partial_t+\Big(\,1+\sum_{i=1}^\infty b_i(r) t^i\,\Big)^{-\tfrac12}\partial_r.\label{prek}\ee
Since the postulated propagation vector $K$ of the wave is spherically symmetric, our ansatz for the metric $\hat{g}$ of the wave spacetime $\hat{M}$ is now spherically symmetric.

\begin{remark}
It is worthwhile to note that, regardless if the metric $\hat{g}$ satisfies Einstein's equations \eqref{eincheck} or not, the \emph{vector} $K$ \emph{is} always \emph{tangent to a congruence of null geodesics without shear and twist}. This represents light rays emanating from the source at the hypersurface $r=0$. We require that the Poincar\'e-Einstein type metric \eqref{prem} satisfies the Einstein equations \eqref{eincheck} with this null vector field $K$ and some functions $\hat{\Phi}$ and $\hat{\Lambda}$.
\end{remark}

Let us count the unknowns: we do not know the coefficients $a_i=a_i(r)$ and $b_i=b_i(r)$ for $i=1,2,\dots,$ and we do not know the functions $\hat{\Phi}=\hat{\Phi}(t,r)$ and $\hat{\Lambda}=\hat{\Lambda}(t,r)$.

\begin{remark}
  Note that if the metric \eqref{prem} satisfies the Einstein equations \eqref{eincheck} the Einstein tensor $\hat{G}{}^{ij}=\hat{R}{}^{ij}-\tfrac12\hat{R}\hat{g}{}^{ij}$ satisfies
  $$\hat{G}{}^{ij}=-\hat{\Lambda}\hat{g}{}^{ij}+\hat{\Phi}K^i K^j.$$
  Thus, a'priori, the symbol $\hat{\Lambda}$ appearing in this equation is a \emph{function} of all the variables, and in general it is \emph{not} a constant. The \emph{effective} energy momentum tensor for the spacetime whose metric satisfies this equation is $\hat{T}{}^{ij}_{\mathrm{eff}}= -\hat{\Lambda}\hat{g}{}^{ij}+\hat{\Phi}K^i K^j$, and the contracted Bianchi identity guarantees that 
  $$\hat{\nabla}_i(-\hat{\Lambda}\hat{g}{}^{ij}+\hat{\Phi}K^i K^j)=0$$
only. Of course this does not imply the constancy of $\hat{\Lambda}$ in general case. This is a \emph{good} feature in a way, because an additional unknown function $\hat{\Lambda}$ prevents the considered Einstein system \eqref{eincheck} to be overdetermined. Having at our disposal this additional unknown will be particularly useful when we will be solving equations \eqref{eincheck} by iterating procedure as explained below. On the other hand it would be desirable that the solution we obtain has constant $\hat{\Lambda}$, since in such a case $\hat{\Lambda}$ could be interpreted as the cosmological constant of $\hat{M}$. Also in such a case, the function $\hat{\Phi}$ would has a nice interpretation as the energy density of the spherical wave described by the physical energy momentum tensor $\hat{T}{}^{ij}=\hat{\Phi}K^i K^j$, which now would be conserved: $\hat{\nabla}_i\hat{T}{}^{ij}=0$. We will see below, that although we do not assume constancy of $\hat{\Lambda}$, our assumption about spherical symmetry, together with Einstein's equations \eqref{eincheck}, are strong enough to actually guarantee that $\hat{\Lambda}$ \emph{is} constant up to an infinite order in $t$. 
  \end{remark}

The strategy of solving the Einstein equations \eqref{eincheck}, for the unknowns $a_i$, $b_i$, $\hat{\Phi}$, $\hat{\Lambda}$, and in turn the spherically symmetric wave spacetime $\hat{g}$ with the wave propagation vector $K$ is as follows:
\begin{itemize}
\item We first calculate the tensor $\hat{E}{}_{ij}:=\hat{R}{}_{ij}-\hat{\Lambda}\hat{g}{}_{ij}-\hat{\Phi}\hat{K}{}_i\hat{K}{}_j$, where $\hat{K}{}_i=\hat{g}{}_{ij}K^j$, with $\hat{g}_{ij}$ as in \eqref{prem}, and $K^i$ as in \eqref{prek}. This in coordinate basis $(x^1,x^2,x^3,x^4)=(z,\bar{z},r,t)$, modulo the symmetry, has the following nonvanishing components: $\hat{E}{}_{12}$, $\hat{E}{}_{33}$, $\hat{E}{}_{34}$ and $\hat{E}{}_{44}$. We use the Einstein equations $\hat{E}{}_{12}=0$ and $\hat{E}{}_{34}=0$ to solve for $\hat{\Lambda}$ and $\hat{\Phi}$. After this move the functions $\hat{\Lambda}$ and $\hat{\Phi}$ are explicitly determined in terms of the unknowns $\mu=\mu(t,r)$, $\nu=\nu(t,r)$ and their partial derivatives up to order \emph{two}. Inserting the so obtained $\hat{\Lambda}$ and $\hat{\Phi}$ to the remaining Einstein equations $\hat{E}{}_{33}=0$ and $\hat{E}{}_{44}=0$ we obtain a system of \emph{two} PDEs for the unknowns $\mu=\mu(t,r)$, $\nu=\nu(t,r)$ and their partial derivatives up to order \emph{two}. These two equations in the original coordinates are quite horrible, but introducing a new coordinate $T$ related to $t$ via $t=Tr$, and denoting the derivative with respect to $T$ as a dot over the function, they can be written in the following form \footnote{I thank Robin Graham for observing that one can write these equations as presented here and explaining this to me in \cite{robin}}:
  \be\begin{aligned}
  R_1(\mu,\nu):=&T^2\ddot{\mu}-2T\dot{\mu}-T^2\ddot{\nu}+2T\dot{\nu}+F_1(\mu,\nu)=0\\
  R_2(\mu,\nu):=&T^2\ddot{\mu}+T^2\ddot{\nu}+2T\dot{\nu}+F_2(\mu,\nu)=0\\
  \end{aligned}
  \label{robil}\ee
  where $F_1$ and $F_2$ are nonlinear differential operators such that $F_1(0,0)=F_2(0,0)=0$ and with the following property:

  Let $s\in \bbN$. Let $\mu_0(T,r)$ and $\nu_0(T,r)$ be smooth functions such that $\mu_0(T,r)=\mathcal{O}(T^3)$, $\nu_0(T,r)=\mathcal{O}(T^3)$ and let $\mu_1(T,r)$ and $\nu_1(T,r)$ be arbitrary smooth functions. Then
  \be F_i(\mu_0+T^s\mu_1,\nu_0+T^s\nu_1)=F_i(\mu_0,\nu_0)+\mathcal{O}(T^{s+1}),\quad i=1,2.\label{robila}\ee
  We will return to this property, and the rest of the argument of Robin Graham's letter \cite{robin} in Section \ref{rogr}, when we will comment about the convergence and uniqueness of our solution for the wave. Here we continue in explaining our strategy of solving the remaining two PDEs $R_1=R_2=0$.
\item We solve $R_1=R_2=0$ iteratively, starting with the metric \eqref{prem} with \emph{linear in} $T$ \emph{terms only}. Thus, we first assume that the metric is in the form 
  $$\hat{g}=\frac{2\big(\,1+\tilde{a}_1(r) T\,\big)\der z \der\bar{z}}{T^2(1+\frac{z\bar{z}}{2})^2}\,+\,\frac{1+\tilde{b}_1(r) T}{T^2r^2}\der r^2-\frac{(\der Tr)^2}{T^2r^2}.$$
  The equations $R_1=R_2=0$ at the order $T^{-1}$ give then $\tilde{a}_1=\tilde{b}_1=0$.
\item Then looking at the metric up to \emph{quadratic} terms in $T$ we take 
  $$\hat{g}=\frac{2\big(\,1+\tilde{a}_2(r) T^2\,\big)\der z \der\bar{z}}{T^2(1+\frac{z\bar{z}}{2})^2}\,+\,\frac{1+\tilde{b}_2(r) T^2}{T^2r^2}\der r^2-\frac{(\der Tr)^2}{T^2r^2}.$$
  Now the equations $R_1=R_2=0$ starts at the terms of order $T^0$. Relating this terms to zero, gives now $\tilde{a}_2=\tilde{b}_2=0$.
\item At the third order in $T$ the situation starts to be interesting: We take now
  $$\hat{g}=\frac{2\big(\,1+\tilde{a}_3(r) T^3\,\big)\der z \der\bar{z}}{T^2(1+\frac{z\bar{z}}{2})^2}\,+\,\frac{1+\tilde{b}_3(r) T^3}{T^2r^2}\der r^2-\frac{(\der Tr)^2}{T^2r^2},$$
  and look at the expressions for $R_1$ and $R_2$. Now, they start at the order $T^1$. The equations for $R_1=R_2=0$ at this order give:
  $$2\tilde{a}_3+\tilde{b}_3=0.$$
  This means that to have $R_1=R_2=0$ up to first order in $T$ the metric must read:
  $$ \hat{g}=\frac{2r^2\big(\,1+\frac{f(r)}{r^3}t^3\,\big)\der z \der\bar{z}}{t^2(1+\frac{z\bar{z}}{2})^2}\,+\,\frac{1-2\frac{f(r)}{r^3}t^3}{t^2}\der r^2-\frac{\der t^2}{t^2},$$
  where we have introduced an arbitrary function $f=f(r)=\tilde{a}_3=-\tfrac12\tilde{b}_3$.
\item At the fourth order we take
  $$\hat{g}=\frac{2\big(\,1+f(r) T^3+\tilde{a}_4 T^4\,\big)\der z \der\bar{z}}{T^2(1+\frac{z\bar{z}}{2})^2}\,+\,\frac{1-2f(r) T^3+\tilde{b}_4(r)T^4}{T^2r^2}\der r^2-\frac{(\der Tr)^2}{T^2r^2},$$
  encountering $T^2$ terms as the lowest order terms in $R_1$ and $R_2$. These terms, when equated to zero give:
  $$\tilde{a}_4=-\tfrac34rf'(r)\quad\mathrm{and}\quad\tilde{b}_4=\tfrac34rf'(r).$$
  Thus to solve  $R_1=R_2=0$ up to terms of order in $T^2$ we have to take
  $$ \hat{g}=\frac{2r^2\big(\,1+\frac{f}{r^3}t^3-\tfrac34\frac{f'}{r^3}t^3\,\big)\der z \der\bar{z}}{t^2(1+\frac{z\bar{z}}{2})^2}\,+\,\frac{1-2\frac{f(r)}{r^3}t^3+\tfrac34\frac{f'}{r^3}t^4}{t^2}\der r^2-\frac{\der t^2}{t^2}.$$
\item At each order of iteration we can use the obtained $\mu$ and $\nu$ and its derivatives to calculate the already determined $\hat{\Lambda}$ and $\hat{\Phi}$. For example, at the 4th order of the metric as above, we calculate that
  $\hat{\Lambda}=3+\mathcal{O}(t^5)$ and that $\hat{\Phi}=3\frac{f'}{r^3}t^6+\mathcal{O}(t^7)$. 
  \end{itemize} 

Continuing up to infinite order, and using the analysis of our Section \ref{rogr}, we arrive at the following theorem\footnote{See also our Remarks \ref{robtra}} and \ref{rotra}.\\

%perfect_fluid_expansion_7_sol.nb
\noindent
\emph{Theorem 1.}\\
If the metric 
\be\begin{aligned}
  \hat{g}=&t^{-2}(-\der t^2+h_t)=\\&
  t^{-2}\,\Big(\,-\der t^2\,+\,\frac{2r^2\big(\,1+\nu(t,r)\,\big)\der z \der\bar{z}}{(1+\frac{z\bar{z}}{2})^2}\,+\,\big(\,1+\mu(t,r)\,\big)\der r^2\,\Big)=
  \\&t^{-2}\,\Big(\,-\der t^2\,+\,\frac{2r^2\big(\,1+\sum_{i=1}^\infty a_i(r) t^i\,\big)\der z \der\bar{z}}{(1+\frac{z\bar{z}}{2})^2}\,+\,\big(\,1+\sum_{i=1}^\infty b_i(r) t^i\,\big)\der r^2\,\Big)\end{aligned}\label{solcheck}\ee
satisfies Einstein's equations
\be \hat{E}{}_{ij}:=\hat{R}{}_{ij}-\hat{\Lambda}\hat{g}{}_{ij}-\hat{\Phi}\hat{K}{}_i\hat{K}{}_j=0\label{eiwei}\ee
with
\be K=K^i\partial_{x^i}=\partial_t+\Big(\,1+\sum_{i=1}^\infty b_i(r) t^i\,\Big)^{-\tfrac12}\partial_r,\quad\quad\hat{K}_i=\hat{g}_{ij}K^j, \label{eiwei1}\ee
then we have:
\begin{itemize}
\item The coefficients $a_1(r)$, $a_2(r)$ $b_1(r)$ and $b_2(r)$ identically vanish, $a_1(r)=a_2(r)=b_1(r)=b_2(r)=0$, and the power series expansion of $h_t$ starts at the $t^3$ terms, $h_t=t^3\chi(r)+\mathcal{O}(t^4)$.
\item The metric $\hat{g}$, or what is the same, the power series expansions $\nu(t,r)=\sum_{i=1}^\infty a_i(r) t^i$ and $\mu(t,r)=\sum_{i=1}^\infty b_i(r) t^i$, are totally determined up to infinite order by an arbitrary differentiable function $f=f(r)$.
\item More precisely, the Einstein equations $\hat{E}{}_{ij}=\mathcal{O}(t^{k+1})$ solved up to an order $k$, together with an arbitrary differentiable function $f=f(r)$,  uniquely determine $\nu(t,r)$ and $\mu(t,r)$ up to the order $(k+2)$.
  \item In the lowest order the solution reads:
    $$\nu=\frac{f}{r^3}t^3+\mathcal{O}(t^4)\quad\mathrm{ and}\quad \mu=-\frac{2f}{r^3}t^3+\mathcal{O}(t^4);$$
    The energy function $\hat{\Phi}$ and the cosmological constant $\hat{\Lambda}$ are:
    $$\hat{\Phi}=3\frac{f'}{r^3}t^6+\mathcal{O}(t^7)\quad\mathrm{ and}\quad \hat{\Lambda}=3+\mathcal{O}(t^{4});$$ the Weyl tensor of the solution is
    $$\hat{W}^i{}_{jkl}=\mathcal{O}(t).$$ In particular, the Weyl tensor $\hat{W}^i{}_{jkl}$ vanishes at $t=0$ and $\hat{\Lambda}>0$ there. 
\end{itemize}

\vspace{0.5cm}
%perfect_fluid_expansion_8_sol_final.nb
\noindent
With the use of computers we calculated this solution up to the order $k=10$, finding explicitly $\nu=\sum_{k=3}^{10}a_kt^k$ and $\mu=\sum_{k=3}^{10}b_kt^k$. The formulas are compact enough up to $k=8$ and up to the order $k=8$ they read:
  %perfect_fluid_expansion_10_sol_final.nb
  $$\begin{aligned}
  \nu(t,r)=&f\tfrac{t^3}{r^3}\,-\,\tfrac34f'\tfrac{t^4}{r^4}+\tfrac{1}{10}\big(-2rf'+3r^2f''\big)\tfrac{t^5}{r^5}+\\&\tfrac{1}{24}\big(3f^2-3rf'+3r^2f''-2r^3f^{(3)}\big)\tfrac{t^6}{r^6}+\\
  &\tfrac{r}{280}\big(-24f'-105ff'+24rf''-12r^2f^{(3)}+5r^3f^{(4)}\big)\tfrac{t^7}{r^7}-\\&
  \tfrac{r}{960}\big(60f'+288ff'-150rf'{}^2-60rf''-216rff''+30r^2f^{(3)}-10r^3f^{(4)}+3r^4f^{(5)}\big)\tfrac{t^8}{r^8}
 +\\&\mathcal{O}(\big(\tfrac{t}{r}\big)^{9})
\end{aligned}
$$
\vspace{0.5cm}
  $$\begin{aligned}
  \mu(t,r)=&-2f\tfrac{t^3}{r^3}\,+\,\tfrac34f'\tfrac{t^4}{r^4}-\tfrac{1}{5}f''\tfrac{t^5}{r^5}\,+\,\tfrac{1}{24}\big(39f^2+r^3f^{(3)}\big)\tfrac{t^6}{r^6}\,-\,\tfrac{r}{280}\big(390ff'+2r^3f^{(4)}\big)\tfrac{t^7}{r^7}+\\&
  \tfrac{r}{960}\big(-18ff'+300rf'{}^2+378rff''+r^4f^{(5)}\big)\tfrac{t^8}{r^8}+\mathcal{O}(\big(\tfrac{t}{r}\big)^{9}).
\end{aligned}
  $$
For a solution up to this order we find that:
$$\scriptsize{\begin{aligned}
  \hat{\Phi}\,=\,&3r^3f'\tfrac{t^6}{r^6}\,+\,3r^3\big(f'-rf''\big)\tfrac{t^7}{r^7}\,+\,\tfrac{3r^3}{2}\big(2f'-2rf''+r^2f^{(3)}\big)\tfrac{t^8}{r^8}\,+\\&\tfrac{r^3}{2}\big(6f'+6ff'-6rf''+3r^2f^{(3)}-r^3f^{(4)}\big)\tfrac{t^9}{r^9}+
  \\&\tfrac{r^3}{8}\big(24f'+66ff'-12rf'{}^2-24rf''-30rff''+12r^2f^{(3)}-4r^3f^{(4)}+r^4f^{(5)}\big)\tfrac{t^{10}}{r^{10}}+
  \\&\tfrac{r^3}{40}\big(120f'+522ff'-177rf'{}^2-120rf''-378rff''+93r^2f'f''+60r^2f^{(3)}+90r^2ff^{(3)}-20r^3f^{(4)}+5r^4f^{(5)}-r^5f^{(6)}\big)\tfrac{t^{11}}{r^{11}}+\\&\mathcal{O}(\Big(\tfrac{t}{r}\Big)^{12}),
  \end{aligned}}$$
$$\hat{\Lambda}=3+\mathcal{O}(t^9).$$
I have no patience to type the Weyl tensor components up to high order. It is enough to say that that up to the 4th order in $t$,  modulo a nonzero constant tensor $C^i{}_{jkl}$, it is equal to:
$$\hat{W}^i{}_{jkl}=\Big(\frac{f}{r^2}\frac{t}{r}-\frac{f'}{r}\frac{t^2}{r^2}+\frac{f''}{2}\frac{t^3}{r^3}\Big) C^i{}_{jkl}+\mathcal{O}(\Big(\tfrac{t}{r}\Big)^4).$$

Of course, for the positivity of the energy density $\hat{\Phi}$ close to the surface $\mathscr{I}^+$ of $\hat{M}$ we need
$$f'>0.$$
\begin{remark}
  Note that the arbitrary function $f=f(r)$ first appears as the leading terms in the \emph{Poincare-Einstein type} expansion for $\nu(t,r)$ and $\mu(t,r)$. These leading terms have an \emph{odd} power in $t$, actually both have power three\footnote{We did not assume it! It was implied by the considered Einstein's equations!}. If for our conformally flat class $[h_0]$ we were looking for the more restrictive \emph{Poincare-Einstein} expansion, we would be forced to admit only even powers of $t$ in the expansion. Then we would be forced to put $f\equiv 0$, which would give 
  $$\hat{g}=t^{-2}\,\Big(\,-\der t^2\,+\,\frac{2r^2\der z \der\bar{z}}{(1+\frac{z\bar{z}}{2})^2}\,+\, \der r^2\,\Big)$$
as the Poincare-Einstein metric. This is the deSitter metric with $\hat{\Lambda}=3$ and $\hat{\Phi}=0$. This observation is in accordance with the theory of conformal invariants, which says that in the analytic category the conformal class $[h]$ in dimension 3 determines the Poincare-Einstein metric uniquely. It is worth noticing that in our case of conformally flat $[h_0]$, even more general Einstein equations \eqref{eiwei} than the usual $Ric(g)=\Lambda g$, imposed on the Poincare-Einstein type expansion also give rise to the unique deSitter metric $\hat{g}$, provided that one only admits even powers of $t$ in the expansion.

Admitting both, even and odd, powers of $t$ we got an arbitrariness in the metric $\hat{g}$ given in terms of a free function $f=f(r)$. In our solution for $\hat{g}$ the odd power terms in $t$ start at order 3, Note that this is the same order as in the case of odd terms in the power series expansion in the The Fefferman-Graham Theorem for the Einstein system $\hat{Ric}(\hat{g})=3\hat{g}$ quoted in Section \ref{swi}. Although we have more general Einstein equations imposed on the metric $\hat{g}$ the behavior of odd power solutions is similar to the Fefferman-Graham expansion: the free function $f=f(r)$ in our solution plays the role of traceless divergence free tensor $h$ from Fefferman-Graham Theorem. In Section \ref{rogr} we show that the appearance of the free function at $t^3$ is related to the properties of the \emph{indicial polynomial matrix} of the PDEs we solve to obtain our solution. This polynomial matrix has $s=3$ as one of its indicial roots. And it is this root for which the corresponding indicial polynomial matrix has nonzero kernel. 

The physical interpretation of the arbitrary function $f=f(r)$ is obvious: it describes the radial modulation of the wave traveling along $K^i$.  
  \end{remark}
\begin{remark}\label{robtra}
  We were unable to find a recurrence relation for the functions $a_i(r)$ and $b_i(r)$ for arbitrary $i>10$. We nevertheless claim that such relations do exist  and that the corresponding power series are convergent. One reason - Robin Graham's sketch of a rigorous mathematical proof of this fact - is presented in Section \ref{rogr}. A heuristic reason for these claims is as follows: Our solution for $\hat{g}$ is a pure radiation Einstein metric with cosmological constant, which have a shearfree expanding but nonwisting congruence of null geodesics which is tangent to the wave propagation vector $K^i$. All such solutions of Einstein's equations are known. They belong to the Robinson-Trautman class of solutions described e.g. in Chapter 28.4 of Ref. \cite{kramer}. Our solution is the spherically symmetric solution from this class \cite{vaidya}, and can be written in terms of the Robinson-Trautman coordinates \cite{RT} as:
  \be \hat{g}=\frac{2v^2\der\zeta\der\bar{\zeta}}{(1+\tfrac12\zeta\bar{\zeta})^2}-2\der u\big(\der v+(1-\frac{2m(u)}{v}-\tfrac13\Lambda v^2)\der u\big).\label{cci}\ee
  The trouble is that, because of the appearance of the free function $m=m(u)$ in \eqref{cci}, there is no an easy way of getting the explicit coordinate transformation from the Robinson-Trautman \emph{null} coordinates $(\zeta,\bar{\zeta},v,u) $ to our coordinates $(z,\bar{z},r,t)$. Such transformation would bring $\hat{g}$ as in \eqref{cci} to ours $\hat{g}$ from \eqref{prem} in which all the coefficients $a_i$ and $b_i$ are determined up to infinite order. Anyhow, knowing this transformation or not, the geometric features of our solution with all $a_i$s and $b_i$s determined for $i\to\infty$, show that our $\hat{g}$  must be identified with the Vaidya solution \eqref{cci}. Thus not only the coefficients $a_i$ and $b_i$ in our solution are determined up to infinite order, but also the power series defining our $\hat{g}$ \emph{converges} to $\hat{g}$ given by \eqref{cci}.
\end{remark}
\begin{remark}\label{rotra}
  Note that the last argument above identifying our iterative solution for $\hat{g}$ with the Vaidya's solution is the only justification for the otherwise only `computer assisted' observation that $\hat{\Lambda}$ of our solution is \emph{constant}, $\hat{\Lambda}=3$. The computer calculations to high orders really suggest that it is true to any order, but neither using Fefferman-Graham analysis, nor using the contracted Bianchi identities in some trivial way, show that it is true to infinite order. So the \emph{purist} would say that the result that 
  $$\hat{\Lambda}=3$$
  for our solution is only a conjecture. We however think that the Vaydia identification argument given in the previous Remark is enough to believe that $\hat{\Lambda}=3$ exactly. 
  \end{remark}

\subsection{Sketch of the proof of the convergence of the power series}\label{rogr}
The following sketch of the proof of the convergence and uniqueness of the solution obtained in the Theorem 1 is due to Robin Graham \cite{robin}. We invoke the relevant quote from \cite{robin} below:

``[The Einstein equations \eqref{robil} define] the linear operator
$$I(T\partial_T)\bma \mu\\\nu\ema=\bma T^2\partial^2_T-2T\partial_T&-T^2\partial^2_T+2T\partial_T\\T^2\partial^2_T&T^2\partial^2_T+2T\partial_T\ema\bma\mu\\\nu\ema.$$
It is called the indicial operator. The matrix valued polynomial obtained by replacing $T\partial_T$ by $s$ is the indicial polynomial:
$$I(s)=s\bma (s-3)&-(s-3)\\(s-1)&(s+1)\ema.$$
One has $$I(T\partial_T)(T^sv)=T^sI(s)v+\mathcal{O}(T^{s+1})$$ for any smooth vector function $v(T,r)=\bma \mu_1(T,r)\\\nu_1(T,r)\ema$. So with $\mu_0$, $\nu_0$, $\mu_1$, $\nu_1$ as [in the first item of our strategy], from \eqref{robil}, \eqref{robila} one obtains
\be
\bma R_1(\mu_0+T^s\mu_1,\nu_0+T^s\nu_1)\\
R_2(\mu_0+T^s\mu_1,\nu_0+T^s\nu_1)\ema=\bma R_1(\mu_0,\nu_0)\\R_2(\mu_0,\nu_0)\ema+T^s I(s)\bma\mu_1\\\nu_1\ema+\mathcal{O}(T^{s+1}).\label{robilo}\ee
The values of $s$ for which $I(s)$ is singular are called the indicial roots. Since $\mathrm{det}\big(I(s)\big) = 2s^3(s-3)$, the indicial roots are $s = 0$ with multiplicity 3 and $s = 3$ with multiplicity 1. Since our initial condition prescribes $\mu(0, r) = \nu(0, r) = 0$, we are given the values of the solution at order $T^0$ . So we only need to consider $s \geq 1$. Thus $s = 3$ is the only relevant indicial root. Observe that $\mathrm{ker}\big(I(3)\big)$ is spanned by $\bma -2\\1\ema$.

The solution can now be constructed inductively. Take $s=1$ and apply \eqref{robilo} with $\mu_0=\nu_0=0$. One concludes that $\mu_1=\nu_1=\mathcal{O}(T)$ since $I(1)$ is nonsingular. Now take $s=2$ and apply \eqref{robilo} with $\mu_0=\nu_0=0$ to conclude $\mu_1=\nu_1=\mathcal{O}(T)$ also for $s=2$. However $I(3)$ is singular, so for $s=3$ one has exactly the freedom to take
$$\bma\mu\\\nu\ema=T^3f(r)\bma -2\\1\ema+\mathcal{O}(T^4).$$
Next, take $s=4$ with $\bma\mu_0\\\nu_0\ema=T^3f(r)\bma -2\\1\ema$. Since $I(4)$ is nonsingular the order 4 perturbation is uniquely determined by $\bma R_1(\mu_0,\nu_0)\\R_2(\mu_0,\nu_0)\ema$, which vanishes to order 4 by construction. This continues to all orders: the $\bma \mu_0\\\nu_0\ema$ at step $s+1$ is taken to be the previously determined solution at order $s$, which satisfies $R_1,R_2=\mathcal{O}(T^{s+1})$ by construction. At each order with $s>3$ the solution is uniquely determined since $I(s)$ is nonsingular for all $s>3$. Thus there is a unique formal power series solution given the free order 3 coefficient $f(r)$.

If $f(r)$ is real analytic, the power series converges. This follows from Theorem 2.2 in \cite{bauendi}.''

We end up this section with the following Corollary which summarizes the results in a long Theorem 1. 

\begin{corollary}
The Poincar\'e-Einstein-type metric \eqref{solcheck} can be interpreted as the ending stage of the evolution of the past eon in Penrose's CCC. The eon has a positive cosmological constant $\hat{\Lambda}\simeq 3$, which is filled with a spherically symmetric pure radiation moving along the null congruence generated by the vector field $K$.  
\end{corollary}

\subsection{Using reciprocity for the model of the present eon}
Now, following the Penrose's reciprocal hypothesis procedure, we summarize the properties of the spacetime $\check{M}$ equipped with the metric $\check{g}$ obtained from $\hat{g}$ as in Theorem 1, by the reciprocal change $\check{\Omega}\to-\hat{\Omega}^{-1}=t$. 
In other words, we are now interested in the properties of the metric $\check{g}=t^4\hat{g}$. We have the following theorem.
\vspace{0.5cm}

\noindent
\emph{Theorem 2.}\\
Assume that the metric $\hat{g}$ as in \eqref{solcheck} satisfies the Einstein equations \eqref{eiwei}-\eqref{eiwei1}, $\hat{E}{}_{ij}=0$. Then, the reciprocal metric
$$\begin{aligned}
  \check{g}=&\check{g}{}_{ij}\der x^i\der x^j=\\&t^4\hat{g}=\\
 & t^2\,\Big(\,-\der t^2\,+\,\frac{2r^2\big(\,1+\nu(t,r)\,\big)\der z \der\bar{z}}{(1+\frac{z\bar{z}}{2})^2}\,+\,\big(\,1+\mu(t,r)\,\big)\der r^2\,\Big)=
  \\&t^2\,\Big(\,-\der t^2\,+\,\frac{2r^2\big(\,1+\sum_{i=1}^\infty a_i(r) t^i\,\big)\der z \der\bar{z}}{(1+\frac{z\bar{z}}{2})^2}\,+\,\big(\,1+\sum_{i=1}^\infty b_i(r) t^i\,\big)\der r^2\,\Big)\end{aligned}$$
satisfies the Einstein equations
\be\check{E}{}_{ij}=\check{R}_{ij}-\check{\Phi}\check{K}_i\check{K}_j-\check{\Psi}\check{L}_i\check{L}_j-(\check{\rho}+\check{p})\check{u}_i\check{u}_j-\tfrac12(\check{\rho}-\check{p})\check{g}_{ij}=0.\label{eiwei2}\ee
Here $\check{K}_i$ and $\check{L}_i$ are the null 1-forms corresponding to the pair of {\color{green}out}going-{\color{red}in}going null vector fields $$K=K^i\partial_{x^i}=\partial_t{\color{green}+}\Big(\,1+\sum_{i=1}^\infty b_i(r) t^i\,\Big)^{-\tfrac12}\partial_r\quad\mathrm{ and}\quad L=L^i\partial_{x^i}=\partial_t{\color{red}-}\Big(\,1+\sum_{i=1}^\infty b_i(r) t^i\,\Big)^{-\tfrac12}\partial_r,$$ via $\check{K}_i=\check{g}_{ij}K^j$ and $\check{L}=\check{g}_{ij}L^j$, and the 1-form vector field $\check{u}_i$ corresponds to the future oriented\footnote{Note that now $t<0$ (!)} timelike unit vector field
$$\check{u}=\check{u}^i\partial_{x^i}=-t^{-1}\partial_t,$$
via $\check{u}_i=\check{g}_{ij}\check{u}^j$.

\vspace{0.7cm}
\noindent
Before giving the explicit formulas for the power expansions of functions $\check{\Phi}$, $\check{\Psi}$, $\check{\rho}$ and $\check{p}$ appearing in this theorem, we make the following remark.\\

\begin{remark}
The Einstein equations \eqref{eiwei2} are equations with an energy momentum tensor consisting of radiation propagating with spherical fronts outward (along $K$) and inward (along $L$); it also consists of a perfect fluid co-moving with the present eon's cosmological time $\tau=-\int t\der t$. Each front of the spherical wave present in the past eon that reached the $t=0$ surface in the present eon produces (i) a \emph{spherical outward wave} with energy density $\check{\Phi}$ going along $K$ out of this sphere, (ii) a \emph{spherical inward wave} with energy density $\check{\Psi}$ going along $L$ \emph{towards the center of this sphere}, and (iii) a portion of a \emph{perfect fluid} with energy density $\check{\rho}$ and isotropic pressure $\check{p}$.  
\end{remark}
For the solutions $\nu(t,r)$, $\mu(t,r)$ of the past eon's Einstein's equations \eqref{eiwei}\eqref{eiwei1}, which were given in terms of the power series expansions as $\nu(t,r)=\sum_{i=3}^{k+2} a_i(r) t^i+\mathcal{O}(t^{k+3})$ and $\mu(t,r)=\sum_{i=3}^{k+2} b_i(r) t^i+\mathcal{O}(t^{k+3})$ in Theorem 1, the formulae for the power series expansions of the energy densities $\check{\Phi}$  $\check{\Psi}$, $\check{\rho}$ and the pressure $\check{p}$ are as follows:
%perfect_fluid_expansion_8_sol_final_inverse_correct.nb
$$\begin{aligned}
  \check{\Phi}=&-\frac{9f}{r^3}t^{-3}\,+\,\frac{9f'}{r^3}t^{-2}\,+\,\frac{1}{2r^4}\big(8f'-11rf''\big)t\,+\,\frac{3}{4r^5}\big(5f'-5rf''+3r^2f^{(3)}\big)\,+\\&\frac{9}{40r^6}\big(16f'+5ff'-16rf''+8r^2f^{(3)}-3r^3f^{(4)}\big)t\,+\\&
  \frac{1}{120r^7}\big(420f'+1068ff'-30rf'{}^2-420rf''-384rff''+210r^2f^{(3)}-70r^3f^{(4)}+19r^4f^{(5)}\big)t^2\,+\\&\dots+\mathcal{O}\big(t^{k-3}\big),
\end{aligned}$$
\vspace{0.5cm}
$$\begin{aligned}
  \check{\Psi}=&-\frac{9f}{r^3}t^{-3}\,+\,\frac{6f'}{r^3}t^{-2}\,+\,\frac{1}{2r^4}\big(2f'-5rf''\big)t^{-1}\,+\,\frac{3}{4r^5}\big(f'-rf''+r^2f^{(3)}\big)\,+\\&\frac{1}{40r^6}\big(24f'-75ff'-24rf''+12r^2f^{(3)}-7r^3f^{(4)}\big)t\,+\\&
  \frac{1}{60r^7}\big(30f'+39ff'+75rf'{}^2-30rf''+33rff''+15r^2f^{(3)}-5r^3f^{(4)}+2r^4f^{(5)}\big)t^2\,+\\&\dots+\mathcal{O}\big(t^{k-3}\big),
\end{aligned}$$
\vspace{0.5cm}
$$\begin{aligned}
  \check{\rho}=&3t^{-4}+\frac{18f}{r^3}t^{-1}\,-\,\frac{18f'}{r^3}\,+\,\frac{-6f'+9rf''}{r^4}t-\,\frac{3}{4r^6}\big(9f^2+3rf'-3r^2f''+2r^3f^{(3)}\big)t^2\,+\\&\frac{3}{20r^6}\big(-24f'+105ff'+24rf''-12r^2f^{(3)}+5r^3f^{(4)}\big)t^3\,-\\&
  \frac{1}{20r^7}\big(60f'+96ff'+120rf'{}^2-60rf''+72rff''+30r^2f^{(3)}-10r^3f^{(4)}+3r^4f^{(5)}\big)t^4\,+\\&\dots+\mathcal{O}\big(t^{k-1}\big),
\end{aligned}$$
\vspace{0.5cm}
$$\begin{aligned}
  \check{p}=&t^{-4}+\frac{6f}{r^3}t^{-1}\,+\,\frac{1}{r^4}\big(2f'-rf''\big)t+\,\frac{1}{2r^6}\big(18f^2+3rf'-3r^2f''+r^3f^{(3)}\big)t^2\,-\\&\frac{3}{20r^6}\big(-8f'+45ff'+8rf''-4r^2f^{(3)}+r^3f^{(4)}\big)t^3\,+\\&
  \frac{1}{30r^7}\big(30f'+57ff'+45rf'{}^2-30rf''+39rff''+15r^2f^{(3)}-5r^3f^{(4)}+r^4f^{(5)}\big)t^4\,+\\&\dots+\mathcal{O}\big(t^{k-1}\big).
  \end{aligned}$$
In these formulas all the \emph{dotted} terms are explicitly determined in terms of $f$ and its derivatives (I was lazy, and typed only the terms adapted to the choice $k=6$ in Theorem 1).

The following remarks are in order:\\

\noindent
\begin{remark}~
\newline
\begin{itemize}
\item Note that since in $\check{M}$ the time  $t<0$, then the requirement that the energy densities are positive near the Big Bang hypersurface $t=0$ implies that
  $$f>0$$
  in addition to $f'>0$, which was the requirement we got from the past eon. Indeed, the leading terms in $\check{\Phi}$ and $\check{\Psi}$ are $\check{\Phi}=\check{\Psi}=-\frac{9f}{r^3}t^{-3}$, hence $\check{\Phi}$ and $\check{\Psi}$ are both positive in the regime $t\to 0^-$ provided that $f>0$. Note also that $f>0$ and $f'>0$ are the only conditions needed for the positivity of energy densities, as the leading term in $\check{\rho}$ is $\check{\rho}\simeq 3t^{-4}$, and is positive regardless of the sign of $t$.
\item Remarkably the leading terms in $\check{\rho}$ and $\check{p}$, i.e. the terms with negative powers in $t$, are proportional to each other with the numerical factor \emph{three}. We have
  $$\check{p}=\tfrac13\check{\rho}+\mathcal{O}(t^0).$$
  This means that immediately after the Bang, apart from the matter content of two spherical in-going and outgoing waves in the new eon, there is also a scattered \emph{radiation} there, described by the perfect fluid with $\check{p}=\tfrac13\check{\rho}$.  \item So what the \emph{Penrose-Tod scenario does to the new eon out of a single spherical wave in the past eon}, is it splits this wave into \emph{three portions of radiation: the two spherical waves}, one which is a damped continuation from the previous eon, the other that is focusing in the new eon, as it encountered a mirror at the Bang surface, \emph{and in addition a lump of scattered radiation described by the statistical physics}.
  
  \end{itemize}
\end{remark}

\section{Acknowledgments} This work is inspired by my long term fascination with the approach to differential invariants via curvature invariants of various classes of pseudo-Riemannian conformal structures. This approach is pioneered and developed by Robin Graham. Over the years I had many contacts and discussions with Robin Graham, from which I always learned a lot. Robin was always willing to answer my questions and to give illuminating explanations of his results. Also this paper has a strong imprint of Robin. In particular Section \ref{rogr} of the paper including a proof of the convergence of the power series solution from Theorem 1 is a quote from a letter from him to me \cite{robin}. This puts the main result of the paper on a solid mathematical ground.

Since the pandemic canceled Robin's 65th Birthday Conference, I as an invited speaker for this event, instead of presenting the result prepared for the conference in person, dedicate it to him in this written form.

\end{document}